\documentclass[11pt,twoside]{article}
\usepackage{./asp2014}

\def\uas        {$\mu$as}
\def\kms        {km s$^{-1}$}
\def\Msun       {M$_\odot$}
\def\hho        {H$_2$O}
\def\Ro         {R$_0$}
\def\To         {$\Theta_0$}

\aspSuppressVolSlug
\resetcounters

\begin{document}

\title{ngVLA: Astrometry and Long Baseline Science}
\author{Mark Reid (1), Laurent Loinard (2) and Thomas Maccarone (3)} 
\affil{((1)Harvard-Smithsonian CfA, Cambridge, MA, USA, 
        (2) Centro de Radioastronom\'ia -- UNAM, Morelia, Michoac\'an, Mexico,
        (3) Texas Tech University, Lubbock, TX, USA)}

\paperauthor{Mark J. Reid}{reid@cfa.harvard.edu}{}{Harvard-Smithsonian CfA}{}{Cambridge}{MA}{02138}{USA}
\paperauthor{Laurent Loinard}{l.loinard@crya.unam.mx}{}{Centro de Radioastronom\'ia -- UNAM,}{}{Morelia}{Michoac\'an}{}{Mexico}
\paperauthor{Thomas Maccarone}{Thomas.Maccarone@ttu.edu}{}{Texas Tech University}{}{Lubbock}{TX}{79409}{USA}

\begin{abstract}
Recent advances in VLBI have led to astrometric accuracy exceeding that of the
Gaia mission goals.   This chapter describes some important astrophysical problems 
that can be addressed with sub-milliarcsecond imaging and micro-arcsecond astrometry
using the ngVLA with long baselines.  
\end{abstract}

\section{Introduction}

Astrometry at centimeter to millimeter wavelengths has advanced dramatically
over the last decade.  VLBI techniques have been perfected and relative positional
accuracy of $\approx10$ \uas\ is routinely achieved for compact sources relative
to background quasars (Reid \& Honma 2014).  Indeed, this demonstrated astrometric 
accuracy exceeds
that of the {\it Gaia} mission goal and, since radio waves can penetrate even 
hundreds of magnitudes of visual extinction, provides unique opportunities to fully
explore the Galactic plane and deeply embedded sources in star forming regions.
Long baseline radio interferometry has produced some remarkable results, including
1) detailed mapping of accretion sub-parsec scale disks around supermassive black holes 
in galaxies 
well into the Hubble flow, yielding ``gold standard'' masses for the black holes and 
direct estimates of the Hubble constant independent of all other methods (Gao et al. 2016), 
and 2) measurements of trigonometric parallax to masers in star forming regions across 
large portions of the Milky Way, revealing its true spiral structure, size, and 
rotation curve (Reid et al. 2014).   

Many potential applications of radio astrometry are currently limited by the 
sensitivity of long baseline interferometry with, for example, ten antennas of 
25-m diameter.  The ngVLA with long baselines comparable to size of North America
could provide a sensitivity increase of more than an order of magnitude, which 
translates to the ability to detect $>30$-times more sources 
(for $N \propto S^{-3/2}$), both for astrometric targets and background quasars.  
The greatly increased background source counts project to a decrease in 
target-quasar separation by a factor of $\sqrt{30}$, which directly translates 
to improved astrometric accuracy by the same factor.  Thus, the ngVLA could 
potentially achieve astrometric accuracies of $\approx1$ \uas!  
In this chapter we explore some of the applications of the astounding capabilities
of the ngVLA with long baselines based on our current knowledge of the Universe.   
Of course, by 2030 and later, new and unknown applications are 
almost certain to be important for the advancement of astrophysics.

\section{Gravitational Wave Sources}

The ngVLA has the potential to directly observe the last stages of the inspiral 
of compact binaries, as well as post merger activity, across the Universe.
The merger of two neutron stars, such as recently observed by {\it LIGO}, or
a neutron star and a black hole, can produce emission across the electromagnetic
spectrum.  Even at distances of hundreds of Mpc, the ngVLA could make 
movies of expanding ejecta or jets, since at cosmological source distances,
scales of $\sim1$ pc correspond to $\sim1$ mas (Alexander et al. 2017).  
One can also expect to witness the inspiral of supermassive black holes (SMBHs), 
complementing a future space-based gravitational wave detector like {\it LISA}, 
by providing the only way to achieve {\it resolved} images of electromagnetic 
counterparts.

The International Pulsar Timing Array goal is to detect gravitational waves
by timing an array of about 30 pulsars.  While distances to these pulsars can
be solved for from the timing data, having to solve for these terms introduces
a noise term and parameter correlations.  Obtaining independent pulsar 
distances via trigonometric parallaxes will be possible with long baselines
on the ngVLA, and this could substantially increase the sensitivity of 
gravitational wave detections.  (See also Section 4 for the importance of
accurate distance for the interpretation of the effects of gravitation radiation.)
 
\section{Black Holes}

{\it Fermi} has detected an amazing variety of energetic phenomena, including beams 
of particles of unexpectedly high energy.  Some of the most energetic sources
in the universe are found in AGN.  One of
the most important and long-standing problems in astrophysics is to answer 
how do cosmic accelerators work and what are they accelerating?  VLBA imaging 
of AGNs has revealed that gamma-ray emission comes from jet components 
(Hodgson et al. 2017), and 
not always from the immediate vicinity of SMBHs.  Still, important questions 
remain: Are jets proton-electron or positron-electron plasma?  What powers 
and collimates AGN jets and how do these jets evolve and interact with their 
host galaxy?  Long baseline radio interferometry provides the highest angular 
resolution possible to make movies of these amazing phenomena on scales down 
to the ergospheres of nearby SMBHs.  Note that recent advances in theory and
computation have opened the possibility of fully relativistic, magneto-hydrodynamic
simulations in 3-dimensions.  High sensitivity and resolution observations with 
long baselines on the ngVLA will be critical to guide simulations and
finally arrive at a deep understanding of these fundamental astrophysical
phenomena.  

\hho\ masers in AGN accretion disks have been used to measure accurate
SMBH masses and the Hubble constant.  But they also have the potential to
determine physical properties of these sub-parsec scale disks.   Of particular
interest is mapping the magnetic fields in these disks via spatially resolved
polarization measurements (Zeeman effect).  Currently, observations have
provided upper limits to magnetic field strength (Modjaz et al. 2005), but, with the greatly
increased sensitivity of the ngVLA, detection should be possible.  Indeed, it will 
be the only telescope that can make such resolved measurements in the foreseeable future. 

Stellar mass black holes pose equally fascinating questions.  Do these 
black holes form from supernova (SN) or by direct collapse without explosions?   
What is the origin of black hole spin?  The ngVLA will be able to provide
unique clues that address these questions by measuring parallax and proper motions.  
For example, the long-standing controversy over the distance to Cyg X-1 
(the source of the famous wager between Thorne and Hawking) was recently 
resolved by long baseline radio astrometry; its accurate parallax distance 
of $1.86\pm0.12$ kpc (Reid et al. 2011) allowed a precise determination of the mass 
of the compact object of $15\pm1$ \Msun\ (Orosz et al. 2011), clearly indicating a black hole, 
and was key to determining that the black hole is spinning maximally with a*>0.92 (Gao et al. 2011).  
This binary is too young for accretion to have appreciably 
spun up the black hole, indicating it was born with great spin.  Regarding
its birth, its measured distance and proper motion (peculiar motion of only 20 km/s)
matches that of the Cyg OB3 star-forming cluster, establishing that it
was born in this young cluster.  Since a SN explosion would have disrupted
the region, the black hole probably formed with a quiet, prompt stellar collapse 
(Mirabel \& Rodrigues 2003).
But this is only one source and similar measurements of many more, weaker sources
with the ngVLA are needed to understand their complete demographics and determine
how they formed.  It is important to note that most known black hole X-ray binaries 
are both too far and too extincted for precise Gaia parallaxes.

At the present time, measurements of astrometric wobble of the black hole 
(induced by the secondary star in the binary) 
are marginally possible in one X-ray binary, Cygnus X-1 (Reid et al. 2011).  
With the ngVLA, if it has enough collecting area on long baselines, proper tracing of 
orbits should become possible for a few X-ray binaries, both due to the increased 
sensitivity and the ability to work at higher frequencies, which will allow both better 
angular resolution and more easily manageable systematics due to less extended emission 
from jets.  This is vital, because the biggest uncertainty in black hole 
mass measurements usually comes from the precision of the estimation of the inclination 
angle of the orbit, since the inferred mass scales as ${\rm sin}^3 i$.  
Additional uncertainties come into play from having incorrect inclination angles using 
either the disk continuum method for spin estimation (e.g. Steiner et al. 2017) 
or the reflection method (e.g. Garcia et al., 2013).

Understanding the distribution of masses of black holes and neutron stars is one of the 
few means we have to probe the actual process of supernova explosions.  
At the present time, there appears to be a gap between the lowest mass black holes 
and the highest mass neutron stars, which would imply that whatever instability causes 
supernovae to actually blow up must be relatively rapid (Belczynski et al. 2012), 
but it remains quite possible that the gap is an artifact of the biases in inclination angle 
measurements from ellipsoidal modulations (Kreidberg et al. 2012).
Having even a few systems where direct inclination angle measurements from astrometric wobble 
can be used to calibrate the ellipsoidal modulation of the secondary's light would be vital 
for building samples of well-understood black hole masses and spins.

\section{Fundamental Physics}

Pulsar parallaxes and proper motions precisely locate these stellar remnants
in the Galaxy and provide full phase-space information.  Coupled with
rotation and dispersion measurements, this can be used to model the 
magnetic field and electron density of the Milky Way (Cordes et al. 1991).  Peculiar motions
(after removing Galactic rotation) give direct information on "kicks" received 
at birth by asymmetrical SN explosions.  Additionally, knowing distances
to pulsars allows accurate mass measurements.  Pulsar mass is the key parameter 
to discriminating among competing models for the equation-of-state of material
at the extreme density of neutron stars.  

One interesting problem critically dependent on the Galaxy's fundamental parameters
(eg, \Ro\ \& \To; see the Milky Way section below) involves the interpretation of the 
orbital decay of the Hulse-Taylor binary pulsar system, owing to gravitational 
radiation as predicted by General Relativity.  In 1993, the Nobel Prize
in Physics was awarded for this measurement.  In order to properly interpret
the observed decay rate, one needs to account for the accelerations of the Sun 
($\Theta_0^2/\rm{R}_0$) and the pulsar ($\Theta(R)^2/R$) from their Galactic orbits.  
In 1993, uncertainties in the values of \Ro\ and \To\ limited the Relativity test to
$\pm0.23$\%.  With improved values based on maser parallaxes, 
this uncertainty was reduced by a factor of 3.  Interestingly, with these values, 
there is a $3\sigma$ discrepancy from General Relativity, using the pulsar distance 
of 9.9 kpc assumed in 1993.  This discrepancy would vanish if the pulsar distance 
is 7.2 kpc (Reid et al. 2014).   
An accurate pulsar parallax is critical to test this prediction,
but may require the sensitivity of the ngVLA as the pulsar is weak.  Of course
there are other binaries that can be used to test General Relativity, and distances to these
coupled with accurate models of the Galaxy are needed and could be supplied 
by the ngVLA with long baselines.

Are physical "constants," such as the fine-structure constant and the
proton-to-electron mass ratio, different in the early Universe?  Does the cosmic 
microwave background temperature evolve as predicted with redshift?   
These and other questions (see section on Galaxy Evolution) can be addressed 
by imaging molecular clouds at high redshift in absorption against bright AGNs.  
However, high sensitivity and angular resolution ($\sim1$ mas) are crucial to 
detecting and isolating (resolving) individual pc-scale clouds, since observations 
at lower resolutions blend together clouds that have a wide range of physical conditions
(Sato et al. 2013).

\section{Galaxy Evolution}

\begin{figure}
\epsscale{0.85} 
\plotone{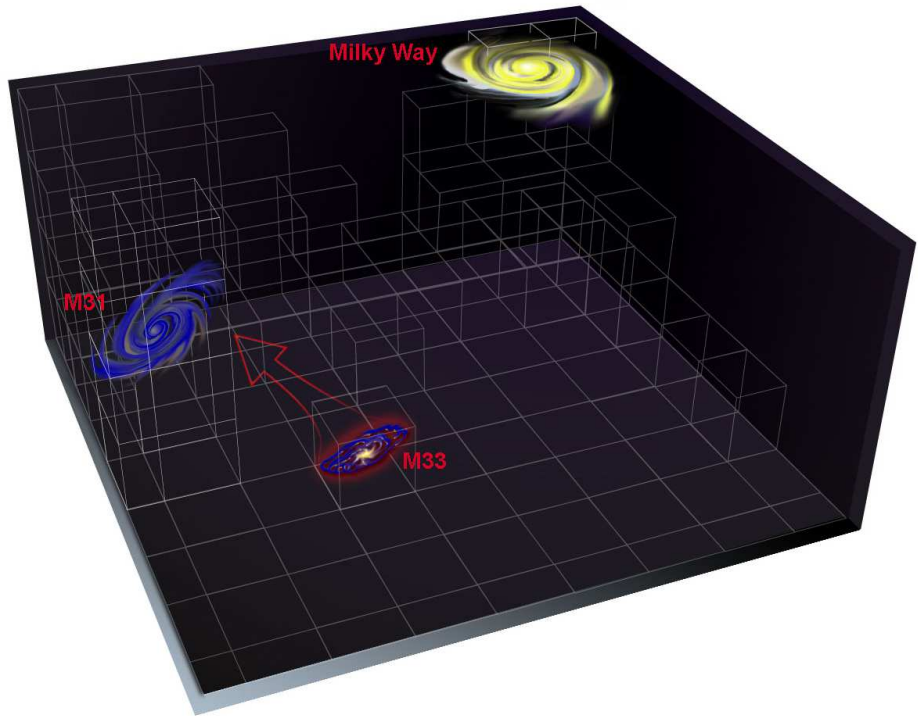}
\caption{\small
Local Group proper motions.  Schematic view of the Local Group of Galaxies with the
Milky Way in the upper right and the Andromeda galaxy (M~31) in the left.   The measured
3-D motion of M~33 is indicated with the red arrow.  Only the radial component, 
approaching the Milky Way at $\approx110$~\kms, of the motion of Andromeda is known.
If Andromeda's proper motion is small it will collide with the Milky Way in a few
billion years.  However, if it has a substantial proper motion ($\sim100$~\kms),
Andromeda and the Milky Way will orbit each other for a very long time. 
 }
\label{fig:local_group}
\end{figure}

The Local Group offers a critical opportunity to study the formation
and evolution of galaxies with extremely high resolution (ie, local cosmology).
Long baseline observations have been able to measure the proper motions of 
two of Andromeda's satellite galaxies, M~33 (Brunthaler et al. 2005) and IC10 
(Brunthaler et al. 2007). Fig.~\ref{fig:local_group} schematically shows the
locations of the Milky Way, M~31 (Andromeda) and M~33, along with the motion
of M~33.  These constrain values of the distance and mass of Andromeda.   
The key parameter for determining the the distribution of dark matter 
and the history and fate of the Local Group is the proper motion of Andromeda itself, 
which would give its 3-D velocity relative to the Milky Way.

Current estimates of Andromeda's proper motion vary considerably, from values
near zero to over 100 km/s.  This range allows scenarios in which Andromeda
directly hits the Milky Way in a few billion years or instead they could orbit
each other for a very long time.  The most direct and accurate measurement
to solve this problem would be astrometric measurements of the AGN in Andromeda, 
M~31*, with \uas\ accuracy over several years.  This may only be possible
with the ngVLA sensitivity coupled with long baselines, 
since M~31* is extremely weak, $\sim10$ $\mu$Jy.

How early do SMBHs form and how do they relate to their protogalaxies?  
When studying protogalaxies, it is crucial to understand what portion of 
the emission is attributable to a SMBH and what comes from a more extended 
star-burst.  At $z\approx6$, a 100 pc sized star-burst nucleus subtends 18 mas 
(and smaller at intermediate redshifts).  For a radio-bright SMBH, the best telescope 
to resolve and separate AGN from star-bursts would be the ngVLA with long baselines.
   
Are molecular clouds seen toward high redshift galaxies different than those 
seen locally?  Molecular absorption line observations at the high angular resolution 
afforded by long baselines on the ngVLA may be the only foreseeable way to probe
the density, temperature, and structure (filamentary?) of molecular clouds
in the early Universe.

\section{Milky Way Structure}

\begin{figure}
\epsscale{0.85} 
\plotone{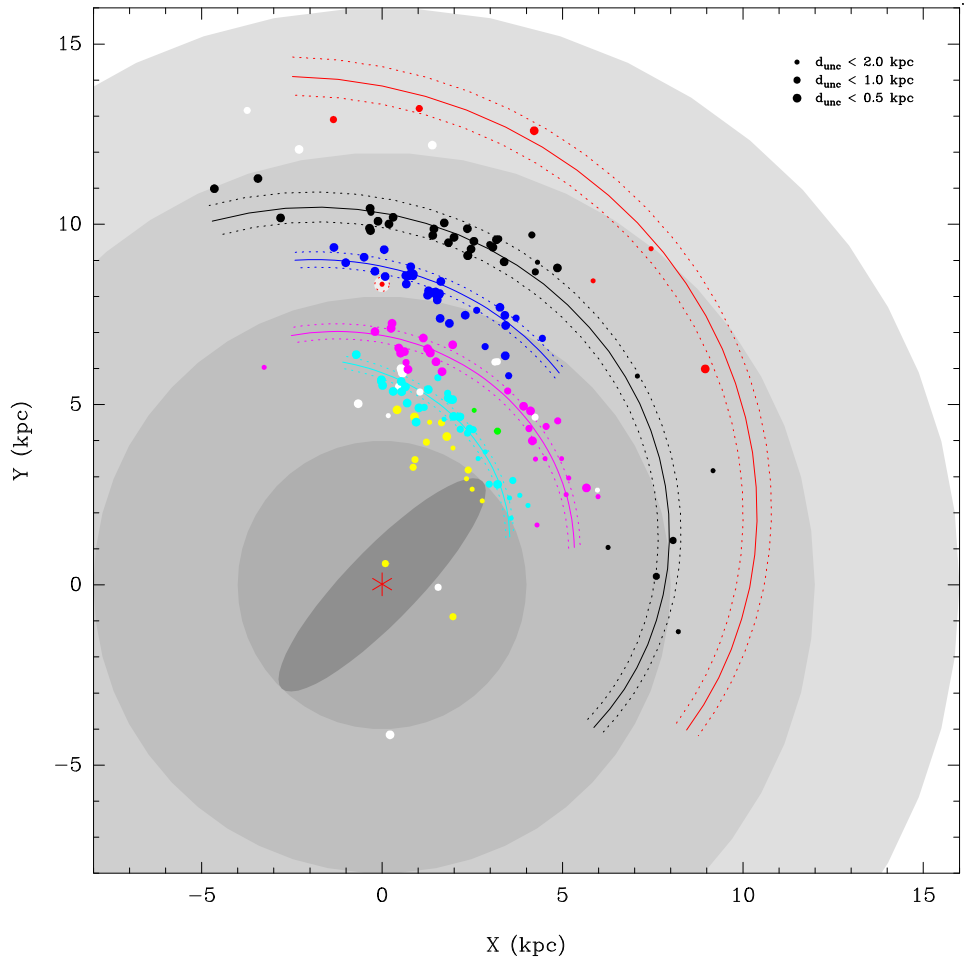}
\caption{\small
Plan view of the Milky Way showing the
locations of high-mass star forming regions (HMSFRs) with 
trigonometric parallaxes measured with VLBI (Reid et al, in preparation).  
The Galactic center ({\it red asterisk}) is at (0,0) and the Sun 
({\it red Sun symbol}) is at (0,8.34).
HMSFRs were assigned to spiral arms based primarily on association with 
structure seen in $\ell-V$ plots of CO and hydrogen emission (and not based on the 
measured parallaxes): Inner Galaxy sources, {\it yellow dots}; Scutum arm, 
{\it cyan octagons};
Sagittarius arm, {\it magenta hexagons}; Local arm, {\it blue pentagons}; 
Perseus arm, {\it black squares}; Outer arm, {\it red triangles}.
Distance errors are indicated in the legend.
The background grey disks provide scale, with radii corresponding in round numbers
to the Galactic bar region ($\approx4$ kpc), the solar circle ($\approx8$ kpc), 
co-rotation of the spiral pattern and Galactic orbits ($\approx12$ kpc), 
and the end of major star formation ($\approx16$ kpc).
The ``long'' bar is indicated with a shaded ellipse.
The {\it solid} curved lines trace the centers (and {\it dotted} lines the $1\sigma$ widths) 
of the spiral arms.
Note that there are few measurements for sources beyond the Galactic center 
(negative Y values); long baselines
with the ngVLA will provide the sensitivity and angular resolution to help finish this map.}
\label{fig:parallaxes}
\end{figure}

VLBI parallax and proper motions of masers have been measured for $\approx200$
high mass star formation regions (HMSFRs), as shown in Fig.~\ref{fig:parallaxes}.  
These clearly trace spiral 
structure across large portions of the Milky Way (Reid et al. 2014), and provide the most
accurate estimates of the distance to the Galactic center, \Ro\ (currently with about 2\% 
accuracy), the rotation speed of the Galaxy at the Sun, \To, and the Galaxy's rotation curve.  
However, with current sensitives, very few sources are detected that are beyond the 
Galactic center.  So, roughly half of the Milky Way remains ``terra incognito.''
The ngVLA with baselines of thousands of km would have the sensitivity
and angular resolution to complete a map of the Milky Way.  The greater sensitivity
of the ngVLA, compared to the current VLBA, allows the use of weaker QSOs that are
then closer in angle to the target sources.  This yields improved parallax accuracy
which coupled with an increased sample size should allow estimation of \Ro\ to 1\% accuracy. 

Pulsar parallaxes can provide an accurate map of their locations in the Galaxy.  
Coupled with rotation and dispersion measurements, they can be used to model the 
magnetic field and electron density of the Milky Way.  Since radio astrometry
improves dramatically at frequencies above a few GHz (to minimize the 
turbulent and difficult to model ionosphere), the greatly improved sensitivity
of the ngVLA, compared to current long baseline interferometers, is needed to do 
astrometry on these very steep-spectrum sources.

\section{Young Stars and Stellar Systems}

The evolution of a star is almost entirely determined by its mass (and, to a much 
smaller extent, its chemical composition). Thus, direct stellar mass measurements 
are of great importance to constrain theoretical stellar evolution models. 
Such measurements can be obtained in multiple stellar systems if the orbital motions 
are monitored with sufficient accuracy and over a period covering 
a significant fraction of an orbital period. 
For main and pre-main sequence stars, this is best achieved using optical and 
near-infrared imaging techniques (adaptive optics, aperture masking interferometry, etc.). 
However, for protostellar objects one must turn to longer wavelengths on account of the 
high extinction that affects such deeply embedded objects. 
Millimeter and centimeter interferometric observations are critical, 
since they can reach both very high angular resolution and high astrometric accuracy. 

At the moment, monitoring of orbital motions has only been possible for a small number of
very young (class 0) stellar systems: eg, IRAS~16293--2422 (Loinard et al. 2002), 
L1551 IRS5 (Rodriguez et al. 2003), and YLW15 (Curiel et al. 2002), via 
multi-epoch VLA observations.  However, none of these observations cover 
a sufficient fraction of an orbit to enable a reliable mass estimate. 
The ngVLA with baselines longer than 300 km and a sensitivity 10-times higher than the 
VLA would permit significant progress in two ways. 
First, relative positions between the 
members of binary systems could be measured to at least 10-times higher accuracy, and 
this would result in a much higher precision in orbit (and, therefore, mass) determination. 
Second, the higher resolution of the ngVLA would enable the 
identification of much tighter systems (with separations as small as $\sim10$ mas, 
compared with $\sim100$ mas for the VLA) with commensurably shorter orbital periods. 
Reliable mass measurements would be attained for such systems. 
It is worth emphasizing that the very early evolution of stars is still a very open 
research topic and that accurate mass measurements at the earliest protostellar stages 
would have a great impact.

Protostellar jets are ubiquitous in young stellar objects, and two mechanisms have been 
proposed to explain their launching.  In magnetospheric models, they are associated with 
the accretion mechanisms that occur on scales of $\sim0.1$ AU ($\sim1$ mas at the 
distance of nearby star-forming regions). For disk-wind models, in contrast, 
the launching occurs on scales comparable with those of the inner accretion disk region 
($\sim10$ AU; or 100 mas). Baselines of at least 300 km would provide an angular 
resolution in the centimeter regime of about 10 mas, which would be sufficient to 
discriminate between the two mechanisms, and to resolve the jets in the transversal 
direction if the disk-wind hypothesis holds. 
However, baselines of at least a few thousand kilometers would be required to resolve 
jets launched through magneto-centripetal acceleration.

\section{Commercial Applications}

While the benefits of the ngVLA to classical astrophysics have been discussed
above, there may be significant other benefits to society.  VLBI observations 
supply accurate telescope locations on a global scale.  This is used to calibrate 
GPS positioning, which is an integral part of our economic activity.   Also VLBI
has demonstrated that it can track spacecraft with exceptional accuracy.
Had VLBI been used to track the Martian lander that crashed into its surface,
the metric-vs-English units error that lead to the multi-billion dollar 
disaster could have been caught and rectified.   
As the USA aims to send robotic crafts and humans
to Mars and beyond, it seems foolish (and dangerous) not to use the best
independent method of spacecraft tracking, which the ngVLA with long baselines
could provide.




\end{document}